\documentclass[12pt]{article}
\usepackage{amsmath}
\usepackage{amssymb}
\usepackage{graphics}
\usepackage{cite}
\usepackage{epsfig}
\usepackage{color}

\newcommand{\beq}{\begin{equation}}
\newcommand{\eeq}{\end{equation}}
\newcommand{\beqa}{\begin{eqnarray}}
\newcommand{\eeqa}{\end{eqnarray}}
\newcommand{\bea}{\begin{eqnarray}}
\newcommand{\eea}{\end{eqnarray}}

\begin{document}

\begin{flushright}
\today
\end{flushright}

\vspace{0.2in}

\begin{center}

\large
{\bf Species Lifetime Distribution}\\
{\bf for Simple Models of Ecologies}

\vspace{0.2in}
\normalsize
{\bf Simone Pigolotti$^{1 *}$, Alessandro Flammini$^{2 *}$,}\\
{\bf Matteo Marsili$^{3 *}$ and Amos Maritan$^{4 *}$}


\normalsize

$^1$Dipartimento di Fisica, Universit\`a di Roma "La Sapienza", \\P.le A.Moro 2,
Rome 00185, Italy\\

$^2$School of Informatics, Indiana University, Bloomington 47408 IN\\

$^3$The Abdus Salam International Center for Theoretical
   Physics, \\Strada Costiera 11, 34100 Trieste, Italy \\

$^4$Dipartimento di Fisica, Universit\`a di Padova, and INFN, via
Marzolo 8, Padova,
 Italy,\\


$^*$All authors have contributed equally to this work.
%

\end{center}

\vspace{0.20in}

\normalsize

\noindent
{\it Corresponding author:} Simone Pigolotti; simone.pigolotti@roma1.infn.it\\

\noindent
{\it Subject Category:} Physical Sciences: Applied Mathematics.


\newpage


\makeatletter
\renewcommand\@biblabel[1]{#1.}
\makeatother

\makeatletter
\makeatother

\begin{abstract}
Interpretation of empirical results based on a taxa's lifetime
distribution shows apparently conflicting results. Species' lifetime
is reported to be exponentially distributed, whereas higher order
taxa, such as families or genera, follow a broader distribution,
compatible with power law decay. We show that both these evidences are
consistent with a simple evolutionary model that does not require
specific assumptions on species interaction. The model provides a
zero-order description of the dynamics of ecological communities and
its species lifetime distribution can be computed exactly.  Different
behaviors are found: an initial $t^{-3/2}$ power law, emerging from a
random walk type of dynamics, which crosses over to a steeper $t^{-2}$
branching process-like regime and finally is cutoff by an exponential
decay which becomes weaker and weaker as the total population
increases. Sampling effects can also be taken into account and shown
to be relevant: if species in the fossil record were sampled according
to the Fisher log-series distribution, lifetime should be distributed
according to a $t^{-1}$ power law.  Such variability of behaviors in a
simple model, combined with the scarcity of data available, cast
serious doubts on the possibility to validate theories of evolution on
the basis of species lifetime data.
\end{abstract}

\section*{Introduction}
Ecosystems have become paradigmatic examples of complex systems,
showing organization and collective dynamics across very different
time and spatial scales \cite{levin}. These features are captured by
non trivial relationships among measurable quantities, which take
forms familiar to statistical physics. Well known examples include
the species-area scaling relationship \cite{harte,durrett},
allometric relations \cite{kleiber,damuth, amos1,amos2}, and the
occurrence of power laws in the distributions of species lifetime
and size of extinction events \cite{sneppen, newman, sole}. These
statistical laws have been measured over many orders of magnitude
and exhibit similar patterns across very different living ecosystems
and also in different quantitative studies of fossil records
\cite{drossel}.  The ubiquity of these patterns \cite{bak} suggests
that they may be amenable to be studied in a general and a-specific
framework.

 In this article, we will address the issue of species (or more
general taxa) lifetime distribution. Although the analysis of fossil
records has recently highlighted several patterns in the evolution of
biodiversity, and motivated the proposition of different mechanisms
that may have caused these patterns, the functional form of the
species lifetime distribution remains a debated issue. According to
several studies \cite{stenseth}, species lifetime seems to be
exponentially distributed. Others have found evidences of power law
behavior with exponent close to -2 if genera, and therefore longer
time scales, are considered (\cite{bak}, see also \cite{drossel} and
reference therein). Keitt and Stanley \cite{stanley} analyzed data
sets from the {\it North American breeding bird survey}
(http://www.mbr.nbs.gov) finding a power law distribution for species
lifetime (in their study defined as the time between colonization and
local extinction) with an exponent close to $-3/2$.  In fact, the
detailed analysis of Newman {\it et al.}  \cite{newsiba} of the data
by Raup \cite{raupsop} has shown how both these hypothesis
consistently fit the data and, when a power law fit is applied, an
exponent between $-3/2$ and $-2$ is estimated.

On the theoretical side, these different, not to say contrasting,
findings have been invoked to support different macro-ecological
theories. The power law behavior with exponent $-2$ is to be expected
when species dynamics can be regarded as a critical branching process
\cite{galton} where two or more species species can originate at a
random moment from a common ancestor and, also randomly, get
extinct. An exponential behavior in the lifetime distribution is often
referred to as Van Valen's law \cite{vanvalen}. The mechanism proposed
by Van Valen in support of this view is commonly known as the Red
Queen effect: there may be no time enough for a species to gain
evolutionary advantage over competing species before the rapidly
changing environment completely re-draws the fitness landscape. As a
consequence, the extinction probability of any species does not depend
on time and an exponential behavior for lifetimes distribution easily
follows. Several data sets support these conclusions
\cite{blow,boersma,saunders,toumarkine} (see also \cite{pearson} for
further analysis of the same data).  More recently, the occurrence of
power law distributions with non trivial exponents has attracted
particular attention, because of an ongoing debate on whether the
observed patterns are caused by a self-organized critical dynamics
\cite{sneppen, newman, sole} that would naturally lead to the notion
of punctuated equilibrium \cite{gould}. In this framework an ecosystem
is depicted as a system of interacting species whose dynamics
converges spontaneously close to a critical point \cite{bak}: the
extinction of a given species may trigger a cascade of extinction
events starting from the species that depend upon, or directly
interact with, the species just extinct and leading to fluctuations of
any size in the number of extinction occurrences that may contiguously
take place.

The aim of the present paper is to show that all the behaviors
mentioned above for the lifetime distribution are captured by a
simple model of non-interacting species. We conclude, therefore, that
it may be problematic, if not inappropriate to discriminate between
existing macroecological theories on the basis of existent datasets.
The framework we adopt here is inspired by the ecological neutral
theory proposed by Hubbell \cite{hubbell} and thereafter extended and
analytically studied in \cite{amos, pigo, volkov,mckane,vallade}.
This class of models assumes that individuals in an ecological
community are fully equivalent and the population of a species is
essentially subject to a birth and death process. Then, each species
undergoes the same dynamics: the reproductive success of each
individuals depends only on the species population size, and not on
the particular species considered. Competition among species is taken
into account explicitly only via a constraint on the total population
of the community and implicitly through averages birth and death
rates.

From the point of view of evolutionary theory, the hypothesis
of species equivalence may be still justified by a Red Queen effect
\cite{vanvalen}, which is able to forbid the acquisition of a large
evolutionary advantage (i.e. a significantly higher fitness level) of
a species over its competitors.  In the framework of population
dynamics, this hypothesis implies that demographic stochasticity is
the main driving force for the assembly of ecological communities,
meaning that its effect is overwhelmingly strong compared to that of
fitness differences among species, which, although present, may be
neglected.  It is worthwhile to stress that, in principle, complex
ecological mechanism acting on long timescales are not ruled out by
these stochastic models, as far as they can be included in effective
birth and death rates. This consideration opens the issue of
determining whether these theories are able to assess realistic
predictions on large time and geographical scales, such as those
relevant for the fossils observations. It is widely believed
\cite{levin} that statistical physics may provide the tools to bridge
these very different scales. In this perspective, it is encouraging
that Conette {\it et al.}, \cite{conette} basing on the studies of the
biodiversity time series compiled by Sepkosky and coworkers
\cite{sopo2}, recently concluded that a random walk-like model is not
inconsistent with the observed biodiversity time-patterns.  

The model we consider here is amenable to be analytically solved and
is introduced in the next section. The resulting lifetime distribution
interpolates, through a scaling function, between the behaviors of two
well known stochastic processes: exit time problem for the one
dimensional random walk \cite{chandra} and the critical Galton-Watson
branching process \cite{galton}. Our results show that, even in a
simple model in which interactions among species are included only in
an averaged way, a variety of different behaviors for the distribution
of extinction time is possible. In particular, depending on the
relevant time-scales, we find an exponential, or a power law
behavior. The latter can either occur with exponent $-2$, typical of
branching processes \cite{galton}, or, for shorter time-scales, with a
random walk like exponent $-3/2$. In addition, if we assume that the
abundance of species is distributed according to a Fisher log-series
\cite{fisher}, in the Galton-Watson case, we find a power law
distribution of extinction times with exponent $-1$.  

As we will discuss in the conclusion, these results stress the
importance of time-scales and sampling effects in the analysis of
lifetime distributions. This theory, also, can easily accommodate the
contrasting empirical observations of Refs.
\cite{stenseth,bak,newsiba,drossel} by assuming that, while species
lifetimes probe the exponential regime of the theory, genera lifetimes
fall in the power law range. The fact that power laws arise in an
``effective`` single-species theory, combined with the sparseness of
available empirical data, suggests that it may not be possible to
validate (or discard) ecological or evolutionary mechanisms like
self-organized critical dynamics \cite{bak} on the basis of an
observed non-exponential behavior in the lifetime distributions.

\section*{Description of the model}

According to the assumption of neutrality \cite{hubbell,
amos}, the dynamics of our model is uniquely specified by the
effective birth and death rates $b^{(n)}$ and $d^{(n)}$ that
depend exclusively on the population size $n$.

We refer to the functions
$b^{(n)}$ and $d^{(n)}$ as "effective" because they may
embody, in a cumulative way, a variety of ecological causes
that may, in principle, influence the increase/decrease over time
of the number of individuals in a species, or, more generally, in
a given taxon. The framework is therefore ample enough to describe
a population dynamics that is not simply dominated by demographic
stochasticity, but also, for example, by immigration, emigration
or niches assembly.  We can safely assume that $b^{(n)}/n$ and
$d^{(n)}/n$, the birth and death rates per individuals, can be
expanded in a power series in $1/n$ around their asymptotic values
$b_1$ and $d_1$ \cite{volkov}:
\begin{eqnarray}\label{bdexp}
b^{(n)}/n=b_1+b_0/n+b_{-1}/n^2+\dots \nonumber\\
d^{(n)}/n=d_1+d_0/n+d_{-1}/n^2+\dots .
\end{eqnarray}

The non-zero coefficient in this Taylor series can be generally related to
various kind of ecological effects giving advantages (or disadvantages) to a
less abundant species with respect to a more abundant one. In
Hubbell's theory \cite{hubbell,amos} the terms $b_0$ and $d_0$
maybe interpreted as the result of an immigration/emigration mechanism which
couples the community to a meta-community living on a larger geographical
scale. The mechanisms described by higher power in $1/n$ in
Eq.(\ref{bdexp}) are relevant only for small population sizes and they
are unable, reasonably, to affect properties observed on large spatial scales and long
timescales. In the following, therefore, we will study
the dynamics described only by the first two terms in the expansion of
Eq. (\ref{bdexp}):
\begin{eqnarray}\label{coefficients}
b^{(n)}=b_0+b_1n\nonumber\\
d^{(n)}=d_0+d_1n
\end{eqnarray}
for all $n \geq 1$. Despite the simple form of the birth and death
rates, and the simplicity of the assumptions, this class of models
is able to provide very good fits of species abundance 
relation \cite{amos, pigo,volkov} which can be related to the
probability $P_n$ of having species with population $n$. This
probability, $P_n$,  evolves with time according to a birth and
death master equation:

\begin{equation}
 \frac{d}{dt} {P}_n(t)=b^{(n-1)}P_{n-1}(t)+d^{(n+1)}P_{n+1}(t)-(d^{(n)}+b^{(n)})P_n(t)
\end{equation}
We impose $b_1<d_1$, ensuring that the average number of
individuals is finite and there is no "demographic explosion". The
ratio $\alpha\equiv b_1/d_1$ fixes, in-fact the average
population per species \cite{pigo,volkov}.

In order to study the lifetime distribution, we consider an
absorbing barrier at $n=0$, imposing $b^{(0)}=d^{(0)}=0$.
The initial condition is that the new species at time $t=0$ has
just one individual:
\begin{equation}\label{initial}
P_n(0)=\delta_{n,1}
\end{equation}
Making these assumptions, $P_0(t)$ represents the probability of
being already extinct at time $t$ and the lifetime probability
distribution function (or exit time distribution), $p(t)$,
is just the time derivative of $P_0(t)$:
\begin{equation}\label{lf}
p(t)= \frac{d}{dt} P_0(t).
\end{equation}
We will first examine the two limit cases $b_1=d_1=0$ and
$b_0=d_0=0$, and then move to the general case.

\section*{Results}
When $b_1=d_1=0$ the number of
individuals belonging to a given species undergoes a random walk
in $n$ space where $b_0$ ($d_0$) is the probability per unit time
to jump one step to the right (left). A species lifetime would
therefore correspond to the time it takes to the random walk to
reach $n=0$, i.e. to exit the positive axis. The problem of exit
time distribution for a random walk process has been widely
studied in the literature (see, for example, \cite{chandra}). In
particular, it is well known that in the critical case $b_0=d_0$
the lifetime follow a distribution of the form $p(t) \sim
t^{-3/2}$. Indeed it is easy to verify that the solution of
Eq. (3), in the present case, is:
\begin{eqnarray}\label{rw}
P_0(t) = & 1-\exp(-2t)( I_0 (2t) + I_1 (2t) ) \nonumber \\
P_k(t) = & \exp(-2t)( I_{k-1}(2t) - I_{k+1}(2t) ) \,\,\,\,\, k>0 ,
\end{eqnarray}
\noindent where $I_k(z) = \frac{1}{\pi} \int_{0}^{\pi} \exp( z
\cos( \theta) ) \cos(k \theta) d \theta$ are modified Bessel
functions of integer order and the unit of time has been chosen
such that $b_0=d_0=1$.  Since for large $z$, $I_0(z)\varpropto
e^z/\sqrt z$ from Eqs. (\ref{lf}) and (\ref{rw}) it follows that
$p(t) \sim t^{-3/2}$ asymptotically.

Let us now analyze the case $b_0=d_0=0$. This limiting case is
interesting from an ecological point of view because the $d_0$ and
$b_0$ terms happen to be small when one looks on a very large
scale (like on continental scale). The dynamics is equivalent to a
Galton-Watson process in continuous time \cite{galton}: the
asymptotic behavior of the lifetime distribution is a classic
result of the theory of critical branching processes
\cite{harris}.

Also in this case, the birth and death equation can be
analytically solved: defining the characteristic function $
G(x,t)=\sum_{n=0}^\infty P_n(t) x^n$ the birth and death equation
can be transformed in a first-order p.d.e. for the function $G$,
which can be integrated with the characteristics method. In the
following, without loss of generality, we set $d_1=1$ and the
initial condition in Eq. (\ref{initial}) translates in $G(x,0)=x$.
As shown in details in the Supplementary Material the exact solution is:
\begin{equation}\label{gwet}
p(t)= \left(\frac{1-\alpha}{e^{(1-\alpha)t}
-\alpha}\right)^2e^{(1-\alpha)t}
\end{equation}

This distribution has an exponential-like shape when $(d_1-b_1)$ (or
$1-\alpha$) is not too small. On the other hand, when $b_1$ approaches
$d_1$, the distribution has a power law behavior with exponent $-2$
and a characteristic timescale $t^*=\frac{1}{1-\alpha}$. The
distribution $p(t)$ can be casted in a more appealing form by using
the language of critical phenomena in statistical mechanics.
For large t and $t/t^*$ fixed, it follows, from
Eq. (\ref{gwet}), that: 
\begin{equation}\label{sf}
p(t)=\frac{1}{t^2}f(\frac{t}{t^*})
\end{equation}
where $f(x)=[x/(1-e^{-x})]^2 e^{-x}$. Thus plotting $t^2p(t)$
versus $t/t^*$ one get, in the scaling region, a universal curve
where all the model details are absorbed in the characteristic
time scale, $t^*$. When dealing with observational data, an
estimate of $t^*$ can be obtained by the ratio of two consecutive
moments, $\langle t^k \rangle$ ($k \geq 1$), of lifetime p.d.f. .

It is also interesting to investigate the role of the initial
condition on the lifetime p.d.f..  Taking into account an
effective speciation rate, one can show, for the particular case
at hand, that the resulting stationary distribution\cite{amos} is
the celebrated Fisher log series \cite{fisher}:
\begin{equation}\label{fisherlog}
P_n=\mathcal{N}\frac{\alpha^n}{n}
\end{equation}
\noindent  where $n>0$ and $\mathcal{N}$ is a normalization
constant.
Using the result above it is therefore possible to calculate
the expected extinction time of a species that is chosen at random in the
ecosystem. Setting as initial
conditions the characteristic function associated to the
distribution (\ref{fisherlog}), $G(x,0)=\log (1-x\alpha)/\log
(1-\alpha)$, one finds:
\begin{equation}
G(0,t)= \frac{1}{\log(1-\alpha)}
\log\left[\frac{(1-\alpha)e^{(1-\alpha)t}}{e^{(1-\alpha)t}- \alpha}\right]
\end{equation}
In this case, again $p(t)=\partial G(0,t)/\partial t \sim
e^{-t/t^*}$ when $t \gg t^*$ whereas $p(t)\sim t^{-1} $ when $t\ll
t^*$, which means that the critical exponent for the
lifetime p.d.f is now $-1$.

\begin{figure}[h]
\begin{center}
\includegraphics[width=10cm]{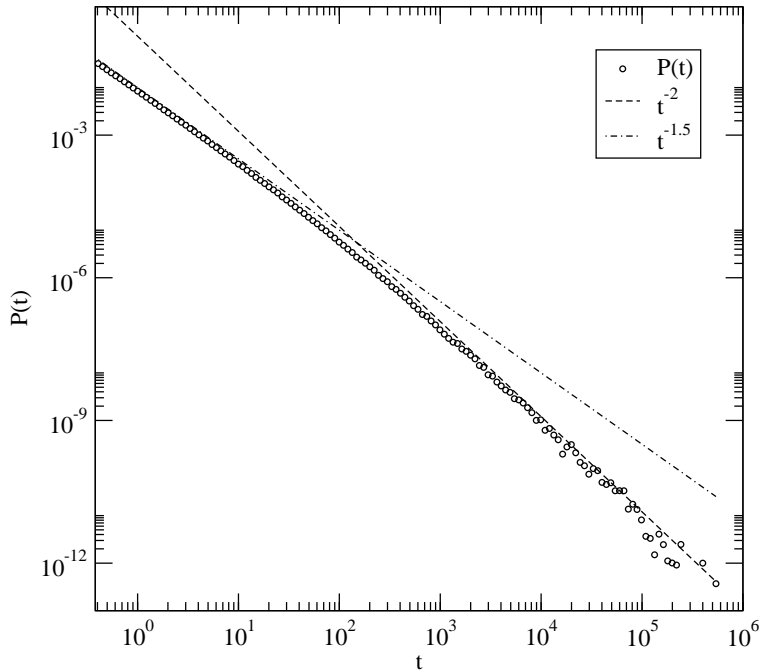}
\end{center}
\caption{numerical sample of the lifetime probability distribution
  function with parameters $d_1=1$, $b_1=1-5\cdot10^{-5}$,
  $b_0=d_0=10$. Notice the crossover between the two power-laws (shown
  in the picture, notice the log-log scale) and the beginning of the
  exponential regime.  }
\label{figure1}
\end{figure}

We now discuss, qualitatively first, the solution in the general
case when all the coefficients are different from zero and $b_0
\sim d_0$, $b_1 \sim d_1$. Heuristically, long-living species have
typically a large number of individuals. For such species the
$b_0$ and $d_0$ terms can be reasonably neglected. Thus, one
expects a crossover from the $t^{-3/2}$ to the $t^{-2}$ behavior
at a certain characteristic time  and finally an exponential decay
beyond another characteristic time scale. Numerical simulations do
support this picture, as shown in Fig.\ref{figure1}, and suggest
that the crossover time is proportional to the ratio
$\frac{b_1}{b_0}$

In the Methods section, we provide the analytical solution of
the general case, proving rigorously both the asymptotic critical
behaviors and the scaling of the solution with the ratio $b_0/b_1$. In
the following, instead, we discuss the main results and their
consequences. In terms of the Laplace transform of $P_0(t)$,
$\tilde{P_0}(s)=\int_0^\infty P_0(t)e^{-ts}$, the exact solution for
the critical case $b_0=d_0=r$ and $b_1=d_1=1$, is given by
\begin{eqnarray}\label{solg}
s \tilde{P}_0(s)-1=-\frac{\int_1^\infty \frac{dy}{y}\
  e^{-sy}(1-\frac{1}{y})^r}{\int_1^\infty dy\ e^{-sy}(1-\frac{1}{y})^r}=
  \frac{1}{\partial_s \log N(s,r)}
\end{eqnarray}
where we have defined $N(s,r)=\int_1^\infty \frac{dy}{y} e^{-sy}
(1-\frac{1}{y})^r$. For small $s$ the function $N(s,r)$ diverges
as $-c\log{s}$, where $c$ depends only on $r$; this implies that
$\tilde{P}_0(s)$ behaves as $\tilde{P}_0(s) \sim \frac{1}{s} +
c\log s$. The Tauberian theorem ensures in this case that $P_0(t)$
behaves like $1-ct^{-1}$  for large $t$, implying that the
lifetime distribution has a $t^{-2}$ power-law tail.

\begin{figure}[h]
\begin{center}
\includegraphics[width=10cm]{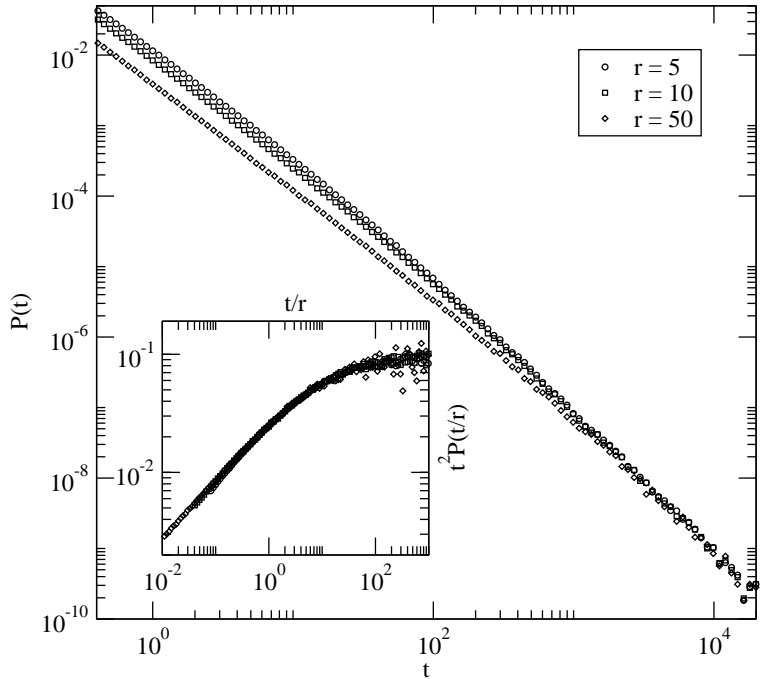}
\end{center}
\caption{Plot of curves with different values of $b_0=d_0=r$, shown in
  the legend. The other parameters are $d_1=1$,
  $b_1=1-10^{-5}$. Collapse of the curves according to the scaling law
  (\ref{scalinglaw}) is shown in the inset.  } \label{figure2}
\end{figure}

In order to derive the crossover to the $t^{-3/2}$ behavior, we need to focus
on time-scales $t\ll 1/r$ for $r \ll 1$. This is related to the limit
$s \to 0$ with $rs$ fixed in the solution, for which one obtains 

\begin{equation}
N(s,r)=\int_{0}^\infty \frac{dx}{x}
e^{-\sqrt{rs}\left(x+\frac{1}{x}\right)}=2K_0(2\sqrt{rs})
\end{equation}
where $K_0$ is a modified Bessel function. Using this result and
Eq. (\ref{solg}) one gets that the lifetime distribution obeys the
following scaling form:

\begin{equation}\label{scalinglaw}
p(t)=t^{-2}f\left(\frac{t}{r}\right)
\end{equation}
where  $f(x)\rightarrow  {\mathrm const}$
when $x \rightarrow \infty$, leading to the $t^{-2}$ scaling at
large $t$, and $f(x) \sim \sqrt{x}$ when $x \rightarrow 0$,
corresponding to the random walk scaling $t^{-3/2}$ at
intermediate $t$. The validity of this scaling law is numerically
confirmed (see Fig.\ref{figure2}).

\section*{Discussion}
 
As sketched in the Introduction, the fact that species lifetimes are
usually exponentially distributed is often referred to as Van
Valen's law \cite{vanvalen}: under the assumption that the fitness
level is correlated in some way to the extinction probability, Van
Valen states that an observed exponential lifetime distribution is the
fingerprint of an acting Red Queen mechanism. Later, more
detailed analysis (and datasets) \cite{newsiba} brought to the
observation of power-law behaviors in genera lifetimes, while
species exponential lifetime distribution have been generally
confirmed.  This difference is, to a certain degree,
counterintuitive as one would expect to see a deviation from
criticality as a finite size effect when looking at long time
scales. A possible explanation proposed in Refs. \cite{drossel,bak} is
that at longer timescales, like those relevant for genera
extinction, collective events like mass extinctions play a more
important role. The interdependence of generic taxa in an ecosystem
generates stronger correlations in their probability to survive, and
these correlations, in turn, may originate a power-law behavior in
the lifetime distribution \cite{drossel,bak}.   

We have shown that also in a simple model, in which every species
undergoes an effective {\it independent} dynamics, a critical behavior
for the lifetimes may occur, with an exponent which is compatible with
the observed value. This critical behavior is generated only by
demographic stochasticity, which is known to be a very important
factor in causing species extinction \cite{macarthur}. Interestingly
enough, the hypothesis underlying this model are not so different to
that bringing Van Valen to the explanation of the exponential species
lifetime: our results clearly indicate that the presence of a Red
Queen effect, i.e. the fitness equivalence of all species, do not
ensure an exponential lifetime distribution, as far as one takes into
account the population sizes in an explicit way. In some sense, in
these models the population size acts as a simple ``memory'' of the
evolutionary history of the species.  

It is worthwhile to connect our approach with a model proposed by Raup
\cite{killcurve} as a null model for the survivorship curves of
Phanerozoic genera (the lifetime distribution can be thought as the
derivative of the survivorship curve). This model, fitting rather well
the fossils data, assumes that species constituting the genera have a
constant speciation and extinction rate. Obviously, the resulting
lifetimes distribution is the same that we recover as a limiting case
of our model in Eq.(\ref{gwet}): the only difference is that, in
Raup's case, the branching-like dynamics is applied at the level of
species (not at the level of individuals). This implies that our model
is well compatible with the data from the fossils record, with the
advantage of being grounded on more realistic (and testable)
hypothesis than the assumption of constant species immigration and
speciation rates.  

In our framework, it is also possible to explain
why the critical behavior in the lifetimes is generally observed when
studying higher taxonomic levels. Let us assume that we can neglect
the terms $b_0$ and $d_0$, as far as we are interested in the tail of
the lifetime distribution. By taking the mean value of the
distribution in Eq. (\ref{fisherlog}), the typical population size can
be expressed as:
\begin{equation}\label{popsize}
\langle n \rangle = \frac{\alpha}{(\alpha-1)} \frac{1}{\log(1-\alpha)}
\end{equation}
The r.h.s of Eq. (\ref{popsize}) diverges when $\alpha \rightarrow
1^-$: thus, a choice of the parameters closer to criticality implies
a larger population size. Since genera lump the  individuals of
may species, the effective value of $\alpha$ for a genera should be
closer to $1$ than it is for species.  Therefore, it may not be
possible to observe the power-law in the species lifetime due to the
experimental error bars and the presence of the exponential cutoff
occurring at $t\sim (1-\alpha)^{-1}$ according to eq.(\ref{gwet})
which, depending on its value, might mask both scaling regimes,
i.e. $t^{-3/2}$ and $t^{-2}$ or only the latter.  

Finally, we
demonstrated that, while 'local' birth and death terms, i.e. terms
that are negligible in the large population size limit, are known to
modify the mean species extinction time \cite{shaffer, lande}, they
are unable to affect the long timescale behavior of the lifetime
distribution: the critical behavior of the distribution, in this
class of models, is uniquely determined by the Galton-Watson part of
the dynamics. Given the robustness of this 'criticality' with
respect to modification of the dynamics on small scale, we suggest
the hypothesis that the observed power law could be simply a
consequence of the branching-like structure of single population
dynamics, rather than an effect of the interactions among different
species.

\section*{Methods}

In these notes we outline the main steps of the derivation of the
results.
Let us firstly focus on the limit $b_0=d_0=0$ when the process reduces
to a Galton - Watson branching process. Without loss of generality, we
can set $d_1=1$ and $b_1=\alpha$ in what follows. 
Introducing the characteristic function 
\begin{equation}
G(x,t)=\sum_{n=0}^\infty P_n(t) x^n,
\end{equation}
the birth and death equation can be transformed in a first-order
p.d.e. for $G(x,t)$
\begin{equation}
\partial_t G(x,t) = (\alpha x^2+1-(\alpha+1)x)\partial_x G(x,t).
\end{equation}
This equation can be integrated using, for example, the characteristic
method (see ref. 1). Taking as initial condition $G(x,0)=x$,
which corresponds to Eq. {\bf 4} in the main text, the complete solution is
\begin{equation}
G(x,t)=\frac{(1-x)-(1-\alpha x)e^{(1-\alpha)t}} {\alpha(1-x)-(1-\alpha
x)e^{(1-\alpha)t}},
\end{equation}
from which we obtain
\begin{equation}
P(0,t)=G(0,t)=\frac{1-e^{(1-\alpha)t}}{\alpha-e^{(1-\alpha)t}},
\end{equation}
and, taking the time derivative of this, we derive Eq. {\bf 7} of the
main text.  It is also easy to see that in the scaling limit, i.e. for
$t^*=1/(1-\alpha)\gg 1$ and $t/t^*$ fixed, $p(t)$ can be cast in the
scaling form {\bf 8}.

In order to deal with the general case, we make a Laplace transform
with respect to time of the generating function and define
\begin{equation}
\tilde{G}(x,s)=\int_0^\infty dt e^{-st} G(x,t)=
\int_0^\infty dt e^{-st} \sum_{n=0}^\infty P_n(t)x^n.
\end{equation}
Then the equation of the dynamics becomes
\begin{eqnarray}
&\left[\alpha x^2+1-(\alpha+1)x \right]\partial_x\tilde{G}(x,s)+\nonumber\\
+&\left[b_0x+\frac{d_0}{x}-b_0-d_0-s\right]
[\tilde{G}(x,s)-g_0(s)]=s g_0(s)-x,
\end{eqnarray}
where we defined $g_0(s)=\tilde{G}(0,s)$, which is the Laplace
transform of $P_0(t)$, the function we wish to compute. Defining
$F(x,s)=\tilde{G}(x,s)-g_0(s)$ and using the fact that $g_0(s)$ does not
depend on $x$, we obtain the following equation for $F(x,s)$:
\begin{equation}\label{ode}
\partial_x F(x,s)+ p(x,s) F(x,s)=q(x,s),
\end{equation}
where
\begin{eqnarray}
p(x,s)&=& \left[\frac{d_0}{x}-\frac{b_0-d_0\alpha}{1-\alpha x}
-\frac{s}{(1-\alpha x)(1-x)}\right]\nonumber\\
q(x,s)&=&\frac{sg_0(s)-x}{(1-\alpha x)(1-x)}.
\end{eqnarray}
Eq. {\bf 7} should be solved with the boundary conditions
\begin{eqnarray}
F(1,s)&=&\frac{1}{s}-g_0(s)\label{cond1}\\
F(0,s)&=&0.\label{cond2}
\end{eqnarray}
Due to the presence of singularities at $x=0$ and $x=1$, some care
must be taken when imposing these conditions on the general solution
of Eq. {\bf 7}. Our strategy is that of solving Eq. {\bf 7} with a
modified initial condition (Eq. {\bf 9}) at $x=1-\epsilon$
\begin{equation}\label{cond1mod}
F(1-\epsilon,s)=\frac{1}{s}-g_0(s).
\end{equation}
Then we will impose condition {\bf 10} on the resulting expression,
which leaves us with an equation for
$g_0(s)$. Finally, we shall restore the boundary condition
{\bf 9} by taking the limit $\epsilon\to 0$. Such an
$\epsilon$-``regularization'' procedure allows us to circumvent the
problem of dealing with the singularities at $x=1$ of 
Eq. {\bf 7}. Notice that, as long as $\alpha=b_1\le 1=d_1$, one
has $\lim_{t\rightarrow \infty}P_0(t)=1$, i.e. the probability of
being asymptotically extinct approaches $1$. 

The generic form of the solution of Eq. {\bf 7} with boundary condition
{\bf 11} is
\begin{equation}\label{risolut}
F(x,s)=e^{\int_x^{1-\epsilon}
  dx'p(x',s)}\left[\frac{1}{s}-g_0(s)\right]
-\int_x^{1-\epsilon}dx'
  q(x',s) e^{\int_x^{x'}
  dx'p(x',s)}.
\end{equation}
The resulting expression is rather complex and it will be considered
later on. We shall first specialize to the particular case $b_0=d_0=r$
and $\alpha=1$ discussed in the main text, which describes the crossover
between the two power law regimes, and then the sub-critical case
$\alpha<1$.

For $b_0=d_0=r$ and $\alpha=1$, the coefficients take the simpler form
\begin{eqnarray}
&p(x,s)=\frac{r}{x}-\frac{s}{(1-x)^2}\nonumber\\
&q(x,s)=\frac{sg_0(s)-x}{(1-x)^2}.
\end{eqnarray}
Up to the leading order in $\epsilon$, the solution is
\begin{equation}
F(x,s)=\frac{e^{-\frac{s}{\epsilon}}(g_0(s)-\frac{1}{s})-\int_x^{1-\epsilon}
  dt \frac{sg_0(s)-t}{(1-t)^2}t^r e^{-\frac{s}{1-t}}}{x^r e^{-\frac{s}{1-x}}}.
\end{equation}
Since the denominator diverges when $x \rightarrow 0$, in order to have
$F(0,s)=0$, we have to impose that the numerator should be equal to
zero. After taking the limit  $\epsilon \to 0$, this yields an
equation for $g_0(s)$, which reads
\begin{equation}
\int_1^x
  dt \frac{sg_0(s)-t}{(1-t)^2}t^r e^{-\frac{s}{1-t}}=0.
\end{equation}
Finally, upon making the substitution $\frac{1}{1-t}=y$ and
rearranging terms, we arrive at our main result, Eq. {\bf 11} of the
main text with $N(s,r)$ given by

\begin{equation}\label{N}
N(s,r)=\int_1^\infty \frac{dy}{y} e^{-sy}
(1-\frac{1}{y})^r
.
\end{equation}

For $r$ fixed and $s\ll 1$, the integral in $N(s,r)$ is dominated by
the region $y\sim 1/s$ and hence $N(s,r) \sim -\log s$; the
application of the Tauberian theorem (see ref. 2) finally demonstrate
the $t^{-2}$ asymptotic behavior of the lifetimes. In order to derive
Eq. {\bf 12} of the paper, in the limit $s\ll 1$ with $rs$ fixed, we make
the change of variables $x=\sqrt{\frac{s}{r}}y$ in Eq. {\bf 16} ,
exponentiate the term $(1-1/y)^r$ in the integral and make a power
expansion
\begin{equation}
N(s,r)=\int_{\sqrt{\frac{s}{r}}}^\infty \frac{dx}{x}
e^{-\sqrt{rs}\left(x+\frac{1}{x}-\sqrt{\frac{s}{r}}\frac{1}{x^2}+\dots\right)},
\end{equation}
which, neglecting corrections of order $\sqrt{s/r}$ leads to Eq. {\bf
12} of the main text. When $rs\gg 1$, i.e. for $t\ll r\gg 1$, we can
use the asymptotic expansion for the modified Bessel function, $K_0$
(see Eq.{\bf 12} of the main text) or, more directly, we can estimate the
integral with the saddle point method: the maximum of the argument of
the exponential occurs at $x^*=1$ and, expanding it to second order
around $x^*=1$, we find
\begin{equation}
N(s,r)\approx e^{-2\sqrt{rs}} \int_{\sqrt{\frac{s}{r}}}^\infty
dx e^{-\sqrt{rs}(x-1)^2}\approx  e^{-2\sqrt{rs}}
{(rs)^{-\frac{1}{4}}}.
\end{equation}
Hence
\begin{equation}
s g_0(s)-1=\frac{1}{\partial_s \log
  N(s,r)}=-\frac{1}{\sqrt{\frac{r}{s}}+\frac{1}{4s}},
\end{equation}
which means that for $s\to 0$, $s g_0(s)-1\sim -\sqrt{s}$
corresponding, according to the Tauberian theorem, to the random walk
behavior $P_0(t)\sim 1/\sqrt{t}$.  The fact that the scaling variable
in the derivation above is $rs$, implies that the crossover time
should be proportional to $r$. Indeed using Eqs.{\bf 11} and {\bf 12}
of the main text and the inverse Laplace transform one derives the
scaling form
\begin{equation}\label{randomscaling}
p(t)=\frac{1}{t^2}f\left(\frac{t}{r}\right),
\end{equation}
where the function $f(x) \sim \sqrt{x}$ for small value of the
argument (i.e. when $x\ll 1$) and approaches a constant when $x$
becomes large.

Finally, let us discuss the sub-critical case $b_1<d_1$. Using exactly
the same strategy as for the critical case, we find that the condition
$F(0,s)=0$ leaves us with the following equation:
\begin{equation}
\int_0^1 dt t^{d_0}\ (1-\alpha t)^{b_0/\alpha-d_0-s/(1-\alpha)-1}
(1-t)^{1/(1-\alpha)-1}(sg_0(s)-t)=0.
\end{equation}
Now, we substitute $y=1-t$ and solve for $g_0(s)$
\begin{equation}
sg_0(s) -1 = -\frac{\int_0^1 dy\
  (1-y)^{d_0}[1-\alpha(1-y)]^{b_0/\alpha-d_0-s/(1-\alpha)-1}\
  y^{s/(1-\alpha)}}{\int_0^1 dy\
  (1-y)^{d_0}[1-\alpha(1-y)]^{b_0/\alpha-d_0-s/(1-\alpha)-1}\
  y^{s/(1-\alpha) -1}}.
\end{equation}
The integral on the numerator is finite when $s\to 0$, whereas that on
the denominator has a leading singularity of order 
$(1-\alpha)/s$. This implies that
$sg_0(s)\simeq -A/[1 + st^{*} ]$, with $A$ constant and $t^* \sim 1/(1 - \alpha)$, which
is exactly the Laplace transform of a distribution of the form
\[
p(t)\simeq e^{-t/t^*}.
\]
This confirms both the asymptotic exponential decay of $p(t)$ and the
scaling of the cutoff time $t^*\sim 1/(1-\alpha)$.

\subsection*{Acknowledgments}
We would like to thank Jayanth Banavar, Igor Volkov and
Tommaso Zillio for collaboration on the subject.

\bibliographystyle{pnas}

\end{document}